\documentclass[prd, aps, onecolumn, showpacs, preprintnumbers, amsmath, amssymb, nofootinbib, letterpaper, superscriptaddress]{revtex4}
\usepackage[dvips]{graphicx}
\usepackage{epsf}
\usepackage{amsmath}
\usepackage{amssymb}

\usepackage{graphicx}
\usepackage{dcolumn}
\usepackage{bm}
\def\be{\begin{equation}}
\def\ee{\end{equation}}
\def\bea{\begin{eqnarray}}
\def\eea{\end{eqnarray}}

\newcommand{\beq}{\begin{equation}}
\newcommand{\eeq}{\end{equation}}

\usepackage{color}

\begin{document}


\title{Bouncing Galileon Cosmologies}

\author{Taotao Qiu}
\email{xsjqiu@gmail.com} \affiliation{ Department of Physics,
Chung-Yuan Christian University, Chung-li 320, Taiwan}
\author{Jarah Evslin}
\email{jarah@ihep.ac.cn} \affiliation{ TPCSF, Institute of High
Energy Physics, CAS, P.O. Box 918-4, Beijing 100049, P.R. China}
\author{Yi-Fu Cai}
\email{ycai21@asu.edu} \affiliation{ Department of Physics, Arizona
State University, Tempe, AZ 85287, USA} \affiliation{ TPCSF,
Institute of High Energy Physics, CAS, P.O. Box 918-4, Beijing
100049, P.R. China}
\author{Mingzhe Li}
\email{limz@nju.edu.cn} \affiliation{ Department of Physics, Nanjing
University, Nanjing 210093, P.R. China}
\author{Xinmin Zhang}
\email{xmzhang@ihep.ac.cn} \affiliation{ TPCSF, Institute of High
Energy Physics, CAS, P.O. Box 918-4, Beijing 100049, P.R. China}


\begin{abstract}

\noindent We present nonsingular, homogeneous and isotropic bouncing
solutions of the conformal Galileon model.  We show that such
solutions necessarily begin with a radiation-dominated contracting
phase. This is followed by a quintom scenario in which the
background equation of state crosses the cosmological constant
boundary allowing for a nonsingular bounce which in turn is followed
by Galilean Genesis. We analyze the spectrum of cosmological
perturbations in this background. Our results show that the
fluctuations evolve smoothly and without any pathology, but the
adiabatic modes form a blue tilted spectrum. In order to achieve a
scale-invariant primordial power spectrum as required by current
observations, we introduce a light scalar field coupling to the
Galileon kinetically. We find two couplings which yield a
scale-invariant spectrum, one of which requires a fine tuning of the
initial conditions. This model also predicts a blue tilted spectrum
of gravitational waves stemming from quantum vacuum fluctuations in
the contracting phase.

\end{abstract}

\maketitle

\newcommand{\eq}[2]{\begin{equation}\label{#1}{#2}\end{equation}}

\section{Introduction}

Nonsingular bouncing cosmologies avoid the cosmological ``Big Bang"
singularity of Standard Cosmology and often solve the horizon
problem, thus they have attracted a lot of attention in the 80 years
since their introduction in Ref.~\cite{Tolman:1931zz}. They have
been studied in models motivated by approaches to quantum gravity
such as modified gravity models \cite{Brustein:1997cv,
Cartier:1999vk, Tsujikawa:2002qc, Biswas:2005qr}, Lagrangian
multiplier gravity actions (see e.g. \cite{Mukhanov:1991zn,
Cai:2010zma}), non-relativistic gravitational actions
\cite{Brandenberger:2009yt, Cai:2009in}, brane world scenarios
\cite{Kehagias:1999vr, Shtanov:2002mb}, torsion gravity
\cite{Cai:2011tc}, ``Pre-Big-Bang" \cite{Gasperini:1992em} and
Ekpyrotic \cite{Khoury:2001wf} cosmology (in which case the
conjectured bounces are classically singular) and loop quantum
cosmology \cite{Bojowald:2001xe}. Bouncing cosmologies have also
been conjectured to occur in some string models. For example, it may
be possible to embed the String Gas Cosmology
\cite{Brandenberger:1988aj} in a bouncing universe
\cite{Biswas:2006bs}. Non-singular bounces may also be studied using
effective field theory techniques by introducing matter fields that
violate certain energy conditions, for example non-conventional
fluids \cite{Bozza:2005wn, Peter:2002cn}, double-field quintom
matter \cite{Cai:2007qw, Cai:2008qb}, fields non-minimally coupled
to gravity \cite{Setare:2008qr} and ghost condensates
\cite{Buchbinder:2007ad, Creminelli:2007aq, Lin:2010pf}. A
non-singular bounce may also arise in a universe with an open
spatial curvature term (see e.g. \cite{Martin:2003sf,
Solomons:2001ef}). A specific realization of a quintom bounce was
obtained in the Lee-Wick cosmology \cite{Cai:2008qw}, see also
\cite{Karouby:2011wj} for its instability. Various bouncing models
were reviewed in Ref.~\cite{Novello:2008ra}.

A successful non-singular bouncing cosmology is typically
accompanied by a violation of the null energy condition (NEC) in a
neighborhood of the bounce \cite{Cai:2007qw}. The NEC, which states
that $T_{\mu\nu}n^\mu n^\nu\geq0$ for any null vector $n^\mu$,
implies that the Hubble parameter of a Friedmann-Robertson-Walker
(FRW) universe is always decreasing throughout the cosmic evolution.
However, a non-singular bounce requires that the time derivative of
the Hubble parameter satisfies $\dot{H}>0$ when $H=0$ so that the
universe is able to transit from a contracting phase to an expanding
one. Moreover, in order to link a stable contraction with a normal
thermal expanding history as observed in our universe, the equation
of state (EoS) of the background universe has to cross the
cosmological constant boundary, which corresponds to the so-called
quintom scenario originally discovered in the context of dark energy
phenomenology \cite{Feng:2004ad}. Consistent models realizing
quintom scenarios are notoriously difficult to construct. For
example in models described by a single perfect fluid or a single
scalar with a Lagrangian of k-essence form
\cite{ArmendarizPicon:1999rj}, the cosmological perturbations
encounter a divergence when the NEC is violated \cite{Feng:2004ad,
Vikman:2004dc, Hu:2004kh, Caldwell:2005ai, Zhao:2005vj}. This
statement was explicitly proven in Ref. \cite{Xia:2007km} as a
``No-Go" theorem for dynamical dark energy models. Furthermore in
the absence of higher derivative terms it was demonstrated that any
violation of NEC in a very general class of models necessarily leads
to either instability or superluminal propagation
\cite{Dubovsky:2005xd}.

To realize a viable quintom scenario of cosmological evolution in a
theory of a single scalar field, one needs to introduce
unconventional operators in the Lagrangian. The simplest and also
the first quintom model was constructed by a combination of a
canonical scalar and a ghost \cite{Feng:2004ad}, which has a
property of cosmic duality \cite{Cai:2006dm}. Later on, it was
realized that the EoS of a scalar can cross $-1$ by introducing
higher order derivative terms \cite{Li:2005fm, Aref'eva:2005fu,
Zhang:2006ck}. However in such models, there is a quantum
instability due to an unbounded vacuum state \cite{Carroll:2003st}.
A possible approach to stable violations of the NEC is the ghost
condensation of Ref. \cite{ArkaniHamed:2003uy}, in which the
negative kinetic modes are bounded via a spontaneous Lorentz
symmetry breaking, although it might allow for superluminal
propagation of information in some cases \cite{Dubovsky:2005xd}.
Various theoretical realizations of quintom scenarios and their
implications for early universe physics were reviewed in
Ref.~\cite{Cai:2009zp} (see also \cite{Qiu:2010ux}).

Recently, beginning with  Ref.~\cite{Nicolis:2008in} a class of
models which stably violates the NEC has been studied extensively.
These are models of a single scalar field called the Galileon. They
are local infrared modifications of general relativity, generalizing
an effective field description of the DGP model \cite{Dvali:2000hr}.
Various Galileon configurations which violate NEC have been shown to
be stable, but generically such configurations admit superluminal
propagation \cite{superlum} and perhaps even closed timelike curves
\cite{taotao}. The key feature of these models is that they contain
higher order derivative terms in the action while the equation of
motion remains second-order in order to avoid the appearance of
ghost modes, realizing the idea pioneered by Horndeski thirty years
ago \cite{Horndeski:1974}. The scalar sector of the theory enjoys a
particular realization of either Galilean or conformal Galilean
symmetry, although this is not respected by ghost-free couplings to
gravity \cite{Alexantigal}.  Later on, this model was generalized to
a DBI version \cite{deRham:2010eu}, the K-Mouflage scenario
\cite{Babichev:2009ee}, the supersymmetric Galileon
\cite{Khoury:2011da}, the Kinetic Gravity Braiding models
\cite{Deffayet:2010qz}, the generic Galileon-like action
\cite{Horndeski:1974, Deffayet:2011gz, Gao:2011qe} and others.
Their phenomenology has been studied extensively, for example in
Refs. \cite{Deffayet:2009wt, Chow:2009fm, Silva:2009km,
Gannouji:2010au, DeFelice:2010pv, Kobayashi:2010cm, Padilla:2010ir,
Hinterbichler:2010xn, Burrage:2010cu, Creminelli:2010qf}.

A class of solutions of the conformal Galileon model which
interpolates between a Minkowski initial state and a final big rip
singularity was described in Ref.~\cite{Creminelli:2010ba}.  These
solutions are stable and in fact are dynamical attractors.  The
authors noted that, as the energy diverges, the effective Galilean
description will break down before the big rip singularity and so
this scenario may provide a new and violent alternative to inflation
which solves the initial singularity and horizon problems. They
referred to this scenario as ``Galilean Genesis" and found that its
evolution is characterized by a strong but stable violation of the
NEC yielding a short duration of fast growth of the background
energy density.  After the Galilean Genesis, other degrees of
freedom become dominant and the universe can be reheated through a
defrosting process by virtue of a coupling of matter to the Galileon
field~\cite{Levasseur:2011mw}.

However one does not need to assume that the initial state is finely
tuned to be a Minkowski universe.  In this note we will show that
generic Galilean Genesis cosmologies begin with a
radiation-dominated contraction, followed by a quintom scenario and
a nonsingular bounce.  This bounce is unusual in several respects.
First, the usual motivations for bounces, the singularity and the
horizon problem, do not apply.  The Galilean Genesis scenario
already is nonsingular and solves the horizon problem.  However, the
fact that the bounce leads into a dynamical attractor means that the
analysis of the inhomogeneity problem associated with bouncing
cosmologies will certainly be different, and perhaps our bounce will
be more phenomenologically viable.  In this background, the
cosmological perturbations of scalar type and tensor type evolve
smoothly and controllably, but as in the usual Galilean Genesis
model are unable to form scale-invariant spectra to explain the
current cosmological observations. As in that case, one may
introduce another light scalar coupled to the Galileon, which plays
a role of a curvaton \cite{Mollerach:1989hu, Linde:1996gt,
Enqvist:2001zp, Lyth:2001nq, Moroi:2001ct} in the bounce model
\cite{Cai:2011zx}. By correctly choosing the couplings, it is
possible to produce a nearly scale-invariant primordial power
spectrum. We consider a kinetic coupling of the curvaton to the
Galileon. Our results show that there exist two couplings which
generate a scale-invariant power spectrum. One of these couplings
features stable curvaton dynamics. If one instead uses the second
coupling, the backreaction grows quickly in the contracting phase
and so the model requires a fine-tuned initial condition. Note that,
similar kinetic couplings have been discussed in Refs.
\cite{Antoniadis:1996dj, Rubakov:2009np, Hinterbichler:2011qk,
Libanov:2011zy}.

We begin in Section II with a quick review of the conformal Galileon
model coupled to gravity and its cosmology. In Section III we
analytically determine the asymptotic behavior of a generic
nonsingular bounce solution and we numerically solve for the entire
evolution. Section IV is devoted to a study of the cosmological
perturbations. The analysis of the curvature perturbations shows
that the evolution of these perturbations is pathology-free, even
during the quintom phase. In order to achieve a scale-invariant
primordial power spectrum, we study the curvaton bounce mechanism
with a generalized kinetic coupling to the Galileon field in Section
V. Section VI presents a summary and discussion.

We use the convention $M_{pl} = 1/\sqrt{8\pi G}$ with $G$ Newton's
gravitational constant.  The signature of the metric is chosen to be
$(-,+,+,+)$.

\section{Galileon Cosmology Basics}

For concreteness we consider the simplest model of a conformal
Galileon scalar field $\Pi$ (which in our notation is dimensionless)
minimally coupled to Einstein gravity. The Lagrangian for $\Pi$
consists of an unusual kinetic term coupling of $\Pi$ to itself, a
nonlinear term similar to that of the DGP model, and a square of the
kinetic term which is required by conformal symmetry
\cite{Nicolis:2008in, Creminelli:2010ba}
\begin{eqnarray}\label{action}
 {\cal S} = \int d^4x \sqrt{-g} \bigg[ \frac{R}{16\pi G}+
 F^2e^{2\Pi}(\partial\Pi)^2+\frac{F^3}{M^3}(\partial\Pi)^2\Box\Pi+\frac{F^3}{2M^3}(\partial\Pi)^4
 \bigg]~.
\end{eqnarray}
Here $R$ is the Ricci scalar of the 4-dimensional space-time,  the
coefficient $F$ is the mass scale of the Galileon field, and $M$ is
another mass scale which suppresses the higher derivative operators.
The Lorentz indices are contracted with the metric $g_{\mu\nu}$ and
the box operator is built from covariant derivatives: $\Box \equiv
g^{\mu\nu} \nabla_\mu \nabla_\nu$.

We note that, when the derivatives are large, neither the
gravitational sector \cite{'tHooft:1974bx, Augusto} nor the Galileon
sector \cite{Nicolis:2008in, taotao} is captured by this action.
This action describes a low energy effective field theory.  In our
bouncing cosmologies the energy density shrinks to zero at early
times, but diverges as one approaches the Galilean Genesis.  Thus
the effective field theory description is reliable except within
some neighborhood of the Genesis where, as in
\cite{Levasseur:2011mw}, one may make use of an entropy field to
reheat the universe  so that the potential strong coupling problem
of Galileon cosmology can be avoided\footnote{A generic scenario for
preheating in a bouncing cosmology was discussed in Ref.
\cite{Cai:2011ci}, where it was shown that a period of stochastic
resonance efficiently takes a universe from a primordial era to the
radiation phase.}.

By varying the action with respect to the inverse metric, one
obtains the stress energy tensor
\begin{eqnarray}
 T_{\mu\nu} &=& -F^2 e^{2\Pi} [2\partial_\mu\Pi\partial_\nu\Pi
 -g_{\mu\nu}(\partial\Pi)^2] \nonumber\\
 && -\frac{F^3}{M^3} [2\partial_\mu\Pi\partial_\nu\Pi\Box\Pi
 -\partial_\mu\Pi\partial_\nu(\partial\Pi)^2
 -\partial_\nu\Pi\partial_\mu(\partial\Pi)^2
 +g_{\mu\nu}\partial_\sigma\Pi\partial^\sigma(\partial\Pi)^2]
 \nonumber\\ && -\frac{F^3}{2M^3} (\partial\Pi)^2
 [4\partial_\mu\Pi\partial_\nu\Pi-g_{\mu\nu}(\partial\Pi)^2]~.
\end{eqnarray}
We are interested in the cosmological evolution of a flat FRW
universe
\begin{eqnarray}\label{metric_FRW}
 ds^2=-dt^2+a^2(t)d\vec{x}^2
\end{eqnarray}
in the conformal Galileon model. Substituting the metric
(\ref{metric_FRW}) into Einstein's equations, we obtain the
Friedmann equations:
\begin{eqnarray}
\label{Hsquare} H^2 &=& \frac{8\pi G}{3}\rho~, \\
\label{Hdot} \dot{H} &=& -4\pi G (\rho+P)~,
\end{eqnarray}
where $H$ is the Hubble parameter characterizing the evolution of
the universe. The energy density and the pressure are
\begin{eqnarray}
\label{rho} \rho &=& F^2 [-e^{2\Pi}\dot\Pi^2
+\frac{1}{\bar{H}^2}(\dot\Pi^4 +4H\dot\Pi^3) ]~,\\ \label{pressure}
P &=& F^2 [-e^{2\Pi}\dot\Pi^2 +\frac{1}{3\bar{H}^2}(\dot\Pi^4
-4\dot\Pi^2\ddot\Pi)]~,
\end{eqnarray}
respectively, where we have introduced the constant
\begin{eqnarray}
 \bar{H} \equiv \sqrt{\frac{2M^3}{3F}}~.
\end{eqnarray}

Inserting the energy density (\ref{rho}) into the first Friedmann
equation (\ref{Hsquare}), one easily solves for Hubble's constant
\begin{eqnarray}\label{Hubble}
 H = 2\alpha\dot\Pi^3 \pm \sqrt{-\alpha\bar{H}^2e^{2\Pi}\dot\Pi^2
 +\alpha\dot\Pi^4 +4\alpha^2\dot\Pi^6}~,
\end{eqnarray}
where
\begin{eqnarray}
 \alpha \equiv \frac{F^2}{3M_p^2 \bar{H}^2} =
 \frac{F^3}{2M^3M_p^2}~
\end{eqnarray}
is a constant with dimensions of area. Note that the reality of the
square root of the above expression (\ref{Hubble}) implies that if
$\dot\Pi$ is nonzero, then its magnitude must be greater than the
strictly positive and $\Pi$-dependent value,
\begin{eqnarray}\label{inequality}
 \dot\Pi^2 \geq \frac{1}{8\alpha} (\sqrt{1
 +16\alpha\bar{H}^2e^{2\Pi}}-1)~,
\end{eqnarray}
which means that $\dot\Pi$ can never cross zero and so the Galileon
scalar $\Pi$ is either a constant or monotonic. Therefore the space
of solutions is not connected.  In particular the Galilean Genesis
inhabits the monotonically increasing branch while a Galilean
Apocalypse inhabits a monotonically decreasing branch.  This
clarifies the claim in Ref. \cite{Creminelli:2010ba} which, based on
an analysis of linear perturbations, concluded that initially
shrinking universes may be in the domain of attraction of a Galilean
Genesis. These initially shrinking universes nevertheless have a
monotonically increasing Galileon field. As we are interested in
configurations which lead to a Galilean Genesis, we will only
consider the branch in while $\Pi$ monotonically increasing.

\section{Galileon Bounce Background Solutions}

In this section, we will describe bouncing solutions in Galileon
cosmology. Recall that a nonsingular bounce requires $\dot{H}>0$
when $H=0$ \cite{Cai:2007qw}. In Eq. (\ref{Hubble}), the first term
of the Hubble parameter is positive as $\Pi$ is increasing.
Therefore, a change of sign of the Hubble parameter in Galileon
cosmology can only occur if the second term of (\ref{Hubble}) is
negative.  Therefore the negative branch of Eq. (\ref{Hubble}) is
needed to obtain a bouncing solution.

\subsection{Asymptotic Solution}

Eq. (\ref{Hubble}) of the previous section determines the Hubble
parameter $H$ as a function of the Galileon field $\Pi$ and its
first derivative. According to the Friedmann equation (\ref{Hdot}),
the Hubble parameter is also determined by the energy density and
the pressure of the Galileon field. Therefore one may use
(\ref{Hubble}) to eliminate the Hubble parameter from
Eqs.~(\ref{Hdot}) and (\ref{rho}) thus obtaining a second order
differential equation for the Galileon field $\Pi$ alone
\begin{eqnarray} \label{edm}
 4\dot\Pi^2\ddot\Pi -3\bar{H}^2e^{2\Pi}\dot\Pi^2 +2\dot\Pi^4
 +12\alpha\dot\Pi^6 = 6\dot\Pi^3\sqrt{B}
 +\frac{\dot{B}}{2\alpha\sqrt{B}}~,
\end{eqnarray}
where we have defined $B$ to be
\begin{eqnarray} \label{B}
 B=-\alpha\bar{H}^2e^{2\Pi}\dot\Pi^2 +\alpha\dot\Pi^4
 +4\alpha^2\dot\Pi^6~,
\end{eqnarray}
which is just the quantity in the square root in Eq. (\ref{Hubble}).

At the moment of the bounce $H=0$ and so using Eq.~(\ref{Hubble})
and the positivity of $\dot\Pi$ \beq \dot\Pi=\bar{H} e^\Pi~. \eeq On
the other hand, before the bounce $H<0$ and so (\ref{Hubble})
implies \beq \dot\Pi>\bar{H} e^\Pi~. \label{ineq} \eeq In this
subsection we will be interested in the behavior of $\Pi$ and the
spacetime geometry in the asymptotic past, therefore the condition
(\ref{ineq}) will be satisfied.

This leads us to two possibilities far in the past.  Either
$\dot\Pi$ becomes so much larger that the $\bar{H} e^\Pi$ terms can
be neglected, or else the ratio of these terms tends to a
constant.\footnote{In principle, the ratio may tend to $0$ but the
smaller term may still be important due to a cancelation, however in
this case the equations of motion without the exponential term would
lead to a solution at leading order which either fails or is
ambiguous at successive orders.  We will see that this is not the
case.}  We will now show that the second possibility is
inconsistent.

Imagine that in the asymptotic past the ratio indeed tended to a
constant $c_0$ \beq \dot\Pi=c_0 \bar{H} e^\Pi~.  \label{paradosso}
\eeq We know from Eq.~(\ref{ineq}) that $c_0\geq 1$. We can easily
solve Eq.~(\ref{paradosso}) to find \beq \Pi=-\ln(c_0 \bar{H}
(t_0-t))~,~~~\dot\Pi=\frac{1}{t_0-t} \eeq for some constant of
integration $t_0$.    If $c_0=1$ then (\ref{edm}) yields the
original Galilean Genesis solution, which violates (\ref{ineq}) and
so $c_0>1$.   Therefore the first two terms in the expression
(\ref{B}) for $B$ are both of order $1/t^4$, and since $c_0\neq 1$
they do not cancel.  This implies that the last term on the right
hand side of the equation of motion (\ref{edm}) is of order $1/t^3$,
whereas all other terms are of higher order in $1/t$.  Thus the
equation of motion (\ref{edm}) has no solution as $t\rightarrow
-\infty$, and we have proved a contradiction.  So no such constant
$c_0$ exists, and in the far past \beq \dot\Pi\gg\bar{H} e^\Pi~.
\eeq Now the equation of motion simplifies slightly to \beq
4\dot\Pi^2\ddot\Pi+2\dot\Pi^4(6\alpha
\dot\Pi^2+1)=6\dot\Pi^5\sqrt{4\alpha^2\dot\Pi^2+\alpha}+\frac{12\alpha\dot\Pi^3+2\dot\Pi}{\sqrt{4\alpha^2\dot\Pi^2+\alpha}}\ddot\Pi~.
\label{eomb} \eeq

Inspecting Eq.~(\ref{eomb}) one sees that the dimensionless quantity
$\alpha\dot\Pi^2$ appears frequently.  Therefore the behavior of the
solution depends strongly on whether $\Pi$ grows more rapidly than
the constant scale $1/\sqrt{\alpha}$.  We can again consider three
cases, corresponding to one scale being much larger and to them
being comparable.

\vspace{.5cm}

\noindent {$\bullet$\bf{\ \ Case 1: Fast roll}}

\noindent First let us consider the case in which in the distant
past \beq \alpha\dot\Pi^2\gg 1~. \label{ineq2} \eeq Then the
equation of motion simplifies to \beq
\frac{1}{2}\dot\Pi^4=2\dot\Pi^2\ddot\Pi~.\label{simp} \eeq Clearly
$\dot\Pi=0$ does not satisfy the condition (\ref{ineq2}), therefore
(\ref{simp}) further simplifies to \beq
\frac{1}{2}\dot\Pi^2=2\ddot\Pi~. \eeq This equation implies that in
the far past $\dot\Pi$ is asymptotic to a constant multiplied by
$1/t$ and therefore goes to zero, violating the hypothesis
(\ref{ineq2}). This hypothesis is therefore inconsistent.

\vspace{.5cm}

\noindent {$\bullet$\bf{\ \ Case 2: Linear evolution}}

\noindent The next possibility is that $\alpha\dot\Pi^2$ tends to a
constant, in which case $\dot\Pi$ tends to a constant while
$\ddot\Pi$ tends to zero.  The latter condition implies that only
the second term on the left side and first on the right side of
Eq.~(\ref{eomb}) survives the infinite past limit.  Squaring both
sides, the $\dot\Pi^{12}$ terms cancel.  The positivity of
$\dot\Pi^2$ then implies that the left hand side is greater, and so
this equation cannot be satisfied, and again the hypothesis that
$\dot\Pi$ tends to a constant is inconsistent.

\vspace{.5cm}

\noindent {$\bullet$\bf{\ \ Case 3: Slow roll}}

\noindent This leaves a single case to be considered \beq
\alpha\dot\Pi^2\ll 1~. \label{ineq3} \eeq In this case the dominant
terms in the equation of motion are \beq
2\dot\Pi^4=\frac{2}{\sqrt{\alpha}}\dot\Pi\ddot\Pi~. \eeq Eliminating
the constant solution branch we are left with \beq
\ddot\Pi=\sqrt{\alpha}\dot\Pi^3 \eeq which is solved by \beq
\dot\Pi=\frac{1}{\alpha^{1/4}\sqrt{2(t_0-t)}} \eeq for an arbitrary
time $t_0$.  Integrating this leads to the general asymptotic
solution \beq \Pi=\Pi_0-\frac{\sqrt{2(t_0-t)}}{\alpha^{1/4}}~.
\label{asymptotic} \eeq

Our second order differential equation (\ref{edm}) led to two
constants of integration, $t_0$ and $\Pi_0$.  While choices of the
constant of integration $t_0$ are related by a time-translation
symmetry, choices of $\Pi_0$ are related by a shift-symmetry of
$\Pi$ which is broken by the exponential term.  This symmetry
appears only in the far past where the exponential term is
negligible.   Therefore the constant of integration $\Pi_0$ labels
physically inequivalent solutions.  The above argument implies that
this single parameter family of solutions contains all nonsingular,
homogeneous and isotropic bouncing solutions of this conformal
Galileon model.  We find bouncing solutions for all positive values
of $\Pi_0$.  As $\Pi_0$ becomes increasingly negative the bounces
become increasingly violent and short, for example at $\Pi_0=-35$ we
find that the scale factor of the universe changes by a factor of 2
in a time $.0001$ as measured in the natural units of
Eq.~(\ref{parameter}).  Beyond this threshold, the gradients become
so steep that the problem is no longer amenable to a numerical
treatment, and probably the effective low energy description breaks
down.  In this paper, for concreteness, we will consider the
solution $\Pi_0=0$, commenting on other cases when they differ
qualitatively.  As $\Pi_0$ is subdominant in the far past, where
cosmological perturbations are generated, this arbitrary choice will
have little effect on our main results.

One can now use Eq.~(\ref{Hubble}) to determine the asymptotic
evolution of the space-time.  To leading order as $t$ tends to
$-\infty$, one finds \beq H=\frac{1}{2(t-t_0)}~. \eeq As desired
this is negative, the universe is contracting.  The coefficient
identifies it as a radiation dominated phase, and as we will review
momentarily, numerically we have verified that the equation of state
in the far past of a bounce solution indeed tends to $w=1/3$.

\begin{figure}
\begin{center}
\includegraphics[scale=0.7]{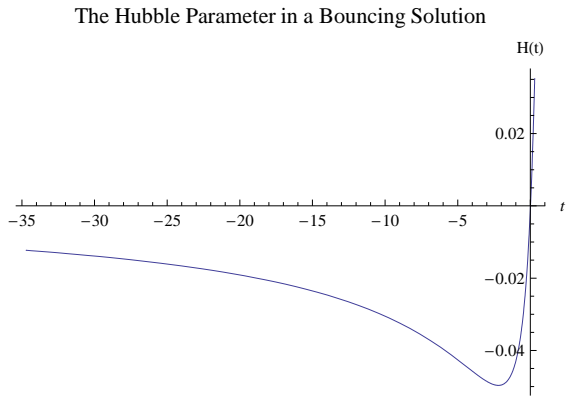}
\caption{The Hubble parameter in a bouncing solution first shrinks
as in a radiation dominated phase, then crosses zero during the
nonsingular bounce and eventually evolves to the moment of Galilean
Genesis. In the numerical calculation, the values of parameters are
listed in (\ref{parameter}). } \label{Hfig}
\end{center}
\end{figure}

\begin{figure}
\begin{center}
\includegraphics[scale=0.7]{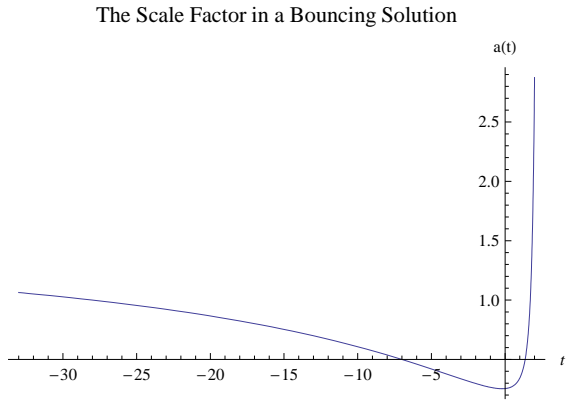}
\caption{The scale factor $a(t)$ in a bouncing solution first
shrinks as in a radiation dominated phase, then arrives at a nonzero
minimal value at the bouncing point and after that enters an
expanding phase. In the numerical calculation, the values of
parameters are listed in (\ref{parameter}). } \label{afig}
\end{center}
\end{figure}

\subsection{Numerical computation}

In the previous subsection we have found that there exists a family
of potential bouncing solutions. However, recall that a bouncing
solution requires that the Hubble constant be initially less than
zero and thus the inequality (\ref{ineq}) places an upper bound on
the Galileon $\Pi$ at any finite value of the cosmic time $t$.
Therefore the existence of a bouncing solution is not guaranteed.
We must rely upon a numerical analysis, whose results we will now
summarize.

\begin{figure}
\begin{center}
\includegraphics[scale=0.7]{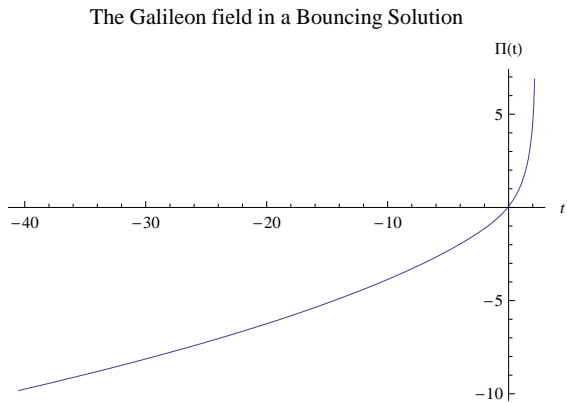}
\caption{The Galileon $\Pi$ monotonically increases throughout the
whole cosmological evolution in the model of Galileon bounce. In the
numerical calculation, the values of parameters are listed in
(\ref{parameter}). } \label{Pifig}
\end{center}
\end{figure}

For concreteness in the numerical calculation, we set
\begin{eqnarray}\label{parameter}
 F = \bar{H} = 1~,~~M_p^2=\frac{1}{8\pi G}=1~,~~\alpha=\frac{1}{3}~,
\end{eqnarray}
and we also make time translation $t\rightarrow t-t{_0}$ in the
numerical plots in order to remove the meaningless constant $t_0$.
We found that the Hubble radius indeed begins as in a radiation
dominated collapse and then becomes positive during the bounce,
ending in a rapidly expanding Galilean Genesis as seen in
Fig.~\ref{Hfig}.  The corresponding scale factor $a(t)$ can be seen
in Fig.~\ref{afig}. The Galileon $\Pi$, as expected, increases
monotonically throughout the evolution as shown in Fig.~\ref{Pifig}.
We can see that $\Pi$ becomes positive near the moment of the
bounce. Therefore, in the expanding phase the factor $e^{2\Pi}$ in
the action is no longer negligible.

\begin{figure}
\begin{center}
\includegraphics[scale=0.7]{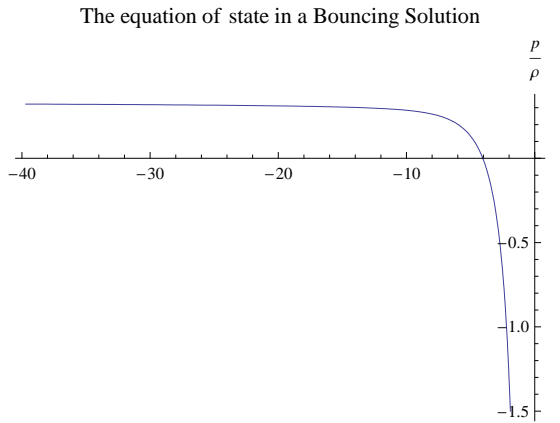}
\caption{The ratio of the pressure to the density of the Galileon
field begins at $1/3$, which is the same as that of normal
radiation. It steadily decreases and crosses $w=-1$ just before the
bounce. In the numerical calculation, the values of parameters are
listed in (\ref{parameter}). } \label{wfig}
\end{center}
\end{figure}

To illustrate that the contracting universe is indeed in a radiation
dominated phase, we plot the evolution of the EoS parameter $w\equiv
P/\rho$, in Fig.~\ref{wfig}. Indeed, we can easily observe that
$w\rightarrow1/3$ in the far past, then it crosses the cosmological
constant boundary $-1$ and falls down toward negative infinity,
which corresponds to the moment of Galilean Genesis. This part of
the numerical computation demonstrates the quintom scenario indeed
is realized in the bounce model of Galileon cosmology.

Before ending this subsection, we would like to make a quick comment
on how possible it will be for a bounce to happen, for different
choice of initial conditions and parameters such as $F$ and $M$.
Since the condition for bounce is to let $H=0$ where $H$ is given by
Eq. (\ref{Hubble}), and we can see that in (\ref{Hubble}) $\Pi$ only
appears in the index of exponential term, so as long as we require
$\Pi$ be negative, this term will be negligible comparing to other
terms, and different value of $\Pi$ will not be sensitive for a
bounce to happen. The initial value of $\dot\Pi$ and the other two
parameters will, however, play an important role in deciding whether
bounce or not, so more or less some fine tuning might be needed.
This could be seen as the price to construct a successful and
healthy bounce model.

\section{Cosmological perturbations and primordial power spectrum}

We have completed our analysis of homogeneous and isotropic bouncing
solutions.  Now we will consider the evolution of cosmological
perturbations on these backgrounds. In particular we will see that
our bouncing solution is stable and free of ghost instabilities
throughout its evolution, even during the quintom and bouncing
phases. At linear order, in standard General Relativity,
cosmological perturbations can be decomposed into scalar and tensor
types \cite{Mukhanov:1990me} which evolve separately. In Galileon
cosmology the equation of motion for the scalar field is still of
second order and the symmetries of our solution and action are still
those of the FRW universe and general relativity respectively.
Therefore we expect that the decomposition of cosmological
perturbations in Galileon cosmology is similar to that in General
Relativity.

A convenient method of studying cosmological perturbation is to work
with the Arnowitt-Deser-Misner (ADM) decomposition of the metric
\cite{Arnowitt:1962hi},
\begin{eqnarray}
 ds^2 = -N^2dt^2 +h_{ij}(dx^i+N^idt)(dx^j+N^jdt)~,
\end{eqnarray}
where $N(x)$ and $N_i(x)$ are the lapse function and the shift
vector, respectively. The tensor $h_{ij}$ is the metric of
3-dimensional space. We can decompose the action (\ref{action}) into
the $3 + 1$ form
\begin{eqnarray}\label{action_ADM}
 {\cal S} = \int dt d^3x \sqrt{h} \frac{N}{2} \bigg[ M_p^2(R^{(3)}
 +K_{ij}K^{ij} -K^2) -4F^2e^{2\Pi}X -\frac{4F^3}{M^3}X\Box\Pi
 +\frac{4F^3}{M^3}X^2 \bigg]~,
\end{eqnarray}
where we have defined
\begin{eqnarray}
 X &\equiv& -\frac{1}{2}(\partial\Pi)^2 \nonumber\\
  &=& \frac{1}{2N^2}(\dot\Pi-N_jh^{ij}\partial_i\Pi)^2
  -\frac{1}{2}h^{ij}\partial_i\Pi\partial_j\Pi~.
\end{eqnarray}
In the above equation, $R^{(3)}$ is the Ricci scalar of a
3-dimensional time-slice. $K_{ij}$ is the corresponding extrinsic
tensor
\begin{eqnarray}
 K_{ij} \equiv \frac{1}{2N}(\dot{h}_{ij}-\nabla_iN_j-\nabla_jN_i)~,
\end{eqnarray}
where $\nabla_i$ is the covariant derivative constructed using the
spatial metric $h_{ij}$ and whose indices are raised and lowered
with $h_{ij}$.

\subsection{Scalar perturbations}

First, we consider scalar cosmological perturbations.  We will work
in the uniform $\Pi$ gauge, in which the perturbations of the
Galileon scalar and the metric are given by
\begin{eqnarray}
 \delta\Pi &=& 0~,\nonumber\\
 h_{ij} &=& a^2e^{2\zeta}\delta_{ij}~.
\end{eqnarray}

Varying the action (\ref{action_ADM}) with respect to the lapse
function $N$ and the shift vector $N_i$ respectively yields the
following Hamiltonian and momentum constraint equations:
\begin{eqnarray}
 \label{constraint1}
 M_p^2(R^{(3)} -K_{ij}K^{ij}+K^{2}) +2F^2e^{2\Pi}\frac{\dot{\Pi}^2}{N^2}
 -\frac{3F^3}{M^3}\frac{\dot\Pi^4}{N^4}
 -\frac{4F^3}{M^3}\frac{\dot\Pi^3}{N^4}\frac{\dot{\sqrt{h}}}{\sqrt{h}}
 +\frac{4F^3}{M^3}\frac{\dot\Pi^3}{N^4}\frac{\sqrt{h}_{,i}}{\sqrt{h}}N^i
 +\frac{4F^3}{M^3}\frac{\dot\Pi^3}{N^4}N^i_{,i} = 0~,
 \\
 \label{constraint2}
 \partial_jK_i^j +\bar{\Gamma}_{jl}^jK_i^l
 -\bar{\Gamma}_{ji}^{l}K_l^j -\partial_iK
 -\frac{2F^3}{M^3M_p^2}\frac{\dot\Pi^3}{N^4}\partial_iN = 0~.
\end{eqnarray}
We expand $N$ and $N_i$ to first order
\begin{eqnarray}
 N = 1 + \varphi ~, ~~ N_i = \partial_i \psi~.
\end{eqnarray}
The constraint equations yield
\begin{eqnarray}
 \varphi
 =\frac{M_{p}^{2}\dot{\zeta}}{M_{p}^{2}H-(F/M)^{3}\dot{\Pi}^{3}}~
\end{eqnarray}
and \begin{eqnarray}
 \psi = \frac{M_{p}^{2}\zeta}{(F/M)^{3} \dot{\Pi}^{3}-M_{p}^{2}H}
 +\frac{a^{2}[3(F/M)^{6}\dot{\Pi}^{6}
 -M_{p}^{2}F^{2}e^{2\Pi}\dot{\Pi}^{2}
 +3(F/M)^{3}M_{p}^{2}\dot{\Pi}^{4}
 +9(F/M)^{3} M_{p}^{2}
 \dot{\Pi}^{3}H]}{[(F/M)^{3}\dot{\Pi}^{3}
 -M_{p}^{2}H]^{2}}\partial^{-2}\dot{\zeta}~.
\end{eqnarray}

Making use of these expressions for $\varphi$ and $\psi$, one can
expand the action up to second order,
\begin{eqnarray}
 {\cal S}_2 = 3\int dt d^3x a^3 D M_p^2 [\dot\zeta^2 -
 \frac{c_s^2}{a^2}(\partial_i\zeta)^2]~,
\end{eqnarray}
where the dimensionless factor $D$ in front of the time derivative
term is
\begin{eqnarray}\label{D_para}
 D = \frac{2M_{p}^{4}H^{2}
 +2(F/M)^{6}\dot{\Pi}^{6}+(F/M)^{3}M_{p}^{2}\dot{\Pi}^{4}}
 {2[M_{p}^{2}H-(F/M)^{3}\dot{\Pi}^{3}]^{2}}~,
\end{eqnarray}
and the sound speed squared is
\begin{eqnarray}\label{cs2_para}
 c_s^2 = \frac{-2M_{p}^{4}\dot{H}
 +2(F/M)^{3}M_{p}^{2}H\dot{\Pi}^{3}-2(F/M)^{6}\dot{\Pi}^{6}
 +6(F/M)^{3}M_{p}^{2}\dot{\Pi}^{2}\ddot{\Pi}}{3[2M_{p}^{4}H^{2}
 +(F/M)^{3}M_{p}^{2}\dot{\Pi}^{4}
 +2(F/M)^{6}\dot{\Pi}^{6}]}~.
\end{eqnarray}

\begin{figure}
\begin{center}
\includegraphics[scale=0.7]{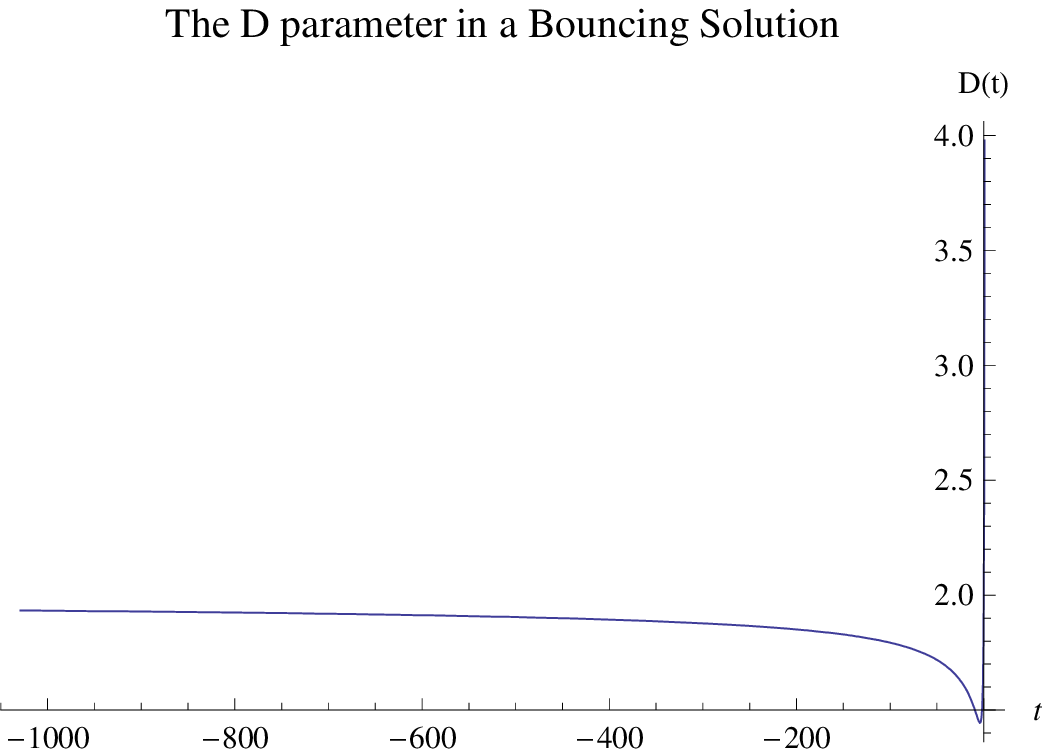}
\caption{Plot of the parameter $D$ in the context of Galileon
bounce. The background parameters are chosen to be the same as the
values provided in Eq. (\ref{parameter}). } \label{Dprofilo}
\end{center}
\end{figure}

\begin{figure}
\begin{center}
\includegraphics[scale=0.7]{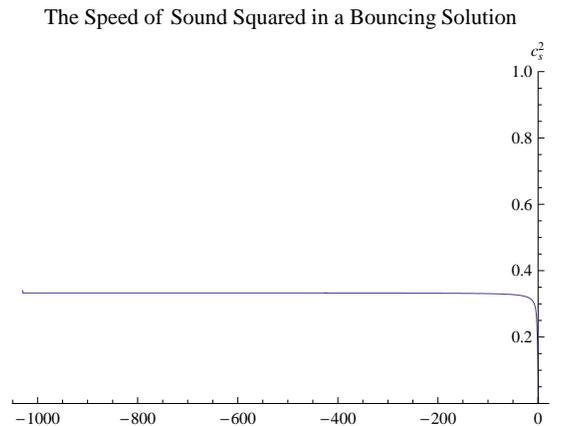}
\caption{Plot of the parameter $c_s^2$ in the context of Galileon
bounce. The background parameters are chosen to be the same as the
values provided in Eq. (\ref{parameter}). } \label{csprofilo}
\end{center}
\end{figure}

We note that a similar but more generic analysis of the perturbation
theory of Galileon cosmology was performed in Ref.
\cite{DeFelice:2011zh} and expressions for $D$ and $c_s^2$ were also
obtained in Ref. \cite{Deffayet:2010qz}. Our results (\ref{D_para})
and (\ref{cs2_para}) are consistent with the corresponding general
calculations in Refs. \cite{DeFelice:2011zh,Deffayet:2010qz}
specialized to the conformal Galileon theory. From Eq.
(\ref{D_para}), we can easily observe that the factor $D$ is
positive definite, and so there do not exist any modes carrying
negative kinetic energy in this background. Consequently, we can
conclude that the Galileon bounce does not suffer from a ghost
instability.

The sound speed $c_s$ characterizes the propagation of a fluctuation
mode with a fixed comoving wavelength and thus determines the
gradient stability. If the numerator of the sound speed squared
(\ref{cs2_para}) is nonnegative, then this model is also be
well-behaved upon the propagation of its perturbation modes.
However, the determination of the sound speed parameter is highly
solution-dependent. Therefore, we will study the propagation of the
scalar perturbation modes in detail.

We numerically plot the evolutions of the parameters $D$ and $c_s^2$
in Figs. \ref{Dprofilo} and \ref{csprofilo}, respectively. From
these two figures, we can see that both parameters are positive
definite throughout the contracting phase and the bouncing phase.
Moreover, the value of $c_s^2$ is less than $1$, which implies there
is no superluminal propagation for the perturbation modes, and so no
risk of closed timelike curves so long as the perturbations remain
small enough to be described by this linear analysis.  At higher
values of $\Pi_0$ there is a bump in $c_s$ around the time of the
bounce, whose height grows with $\Pi_0$, however the maximum
velocity falls just short of the speed of light\footnote{ Indeed in
the high $\Pi_0$ limit there is a long, slow bounce during which the
universe is essentially Minkowski.  In this sense, the usual
Galilean Genesis cosmology can be considered to be the large $\Pi_0$
limit of our scenario.}.  Thus small perturbations about our
background are causal.

\begin{figure}
\begin{center}
\includegraphics[scale=0.7]{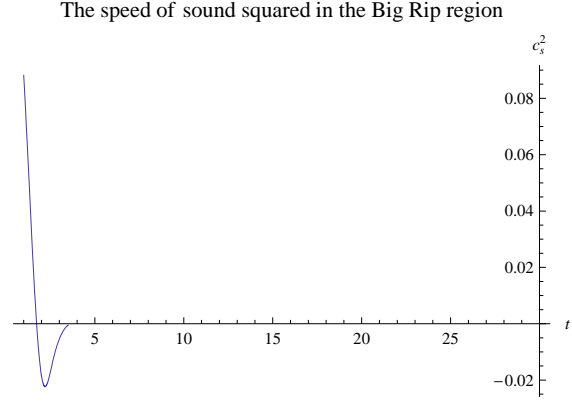}
\caption{Long after the bounce, during the Galilean Genesis, the
speed of sound squared dips below zero.  Naively this indicates a
gradient instability, however the Galileon effective theory cannot
be trusted in the high energy density regime close to the big rip
singularity.  In the numerical calculation, the values of parameters
are listed in (\ref{parameter}). } \label{bad}
\end{center}
\end{figure}

After the bouncing phase, the universe will approach the moment of
Galilean Genesis.  Here $D$ divergences and $c_s^2$ briefly passes
below zero, as seen in Fig.~\ref{bad}.  The negative value of
$c_s^2$ indicates an instability in the model, however as has been
stressed in Ref. \cite{Creminelli:2010ba} the high energy density
implies that perturbative results obtained from the low energy
effective Galileon theory are not to be trusted in this region.  In
that note the authors were unable to characterize the scale at which
the low energy effective Galileon description breaks down, arguing
merely that this occurs before the Planck scale.  Here however the
negativity of $c_s^2$ very close to the Galilean genesis solution
provides a novel upper bound on the latest time at which the
Galileon effective theory is reliable.

We note however that when $\Pi_0\lesssim -3$, $c_s^2$ dips below
zero before the bounce.   This suggests that bouncing solutions are
only stable for sufficiently large initial values of the Galileon
$\Pi$. This is not so surprising, in light of the aforementioned
observation that at very low values of $\Pi_0$ the time derivatives
become very large even before the bounce and so one expects a
premature breakdown of the effective theory in this parameter range.
Therefore we learn that a Galilean genesis can only occur if the
initial condition $\Pi_0$ exceeds a critical value of about $-3$.

In order to provide a well-defined quantization for the cosmological
perturbations, it is usually convenient to redefine the perturbation
variable so that the second order action is of canonical form. In
the case of Galileon cosmology, we introduce the variable
\begin{eqnarray}
 u \equiv z\zeta~, ~~z\equiv a\sqrt{D}~.
\end{eqnarray}
The momentum space equation of motion for a mode of this new
variable is
\begin{eqnarray}
 u_k'' + (c_s^2k^2 - \frac{z''}{z})u_k = 0~,
\end{eqnarray}
where the prime denotes the derivative with respect to the conformal
time  $\eta \equiv \int a^{-1}(t)dt$.

Recall that, from the study of background dynamics, one has learned
that the universe is radiation-dominated during the contracting
phase. Thus it is easy to find $a(\eta)\sim\eta$ and so $a''=0$.
Making use of this relation, one can write  the perturbation
equation in a Klein-Gordon form
\begin{eqnarray}\label{eompert}
 u_k''+(c_s^2k^2+a^2m_{eff}^2)u_k = 0~,
\end{eqnarray}
where the effective mass squared is defined as
\begin{eqnarray}
 m_{eff}^2 =
 \frac{\dot{D}^2}{4D^2}-\frac{\ddot{D}}{2D}-\frac{3H\dot{D}}{2D}~.
\end{eqnarray}

From the background solution (\ref{asymptotic}), we can learn that
in the far past $\dot\Pi \sim \eta^{-1}$ while $H \sim -\eta^{-2}$.
Using this asymptotic solution, we obtain the approximate forms for
the parameters
\begin{eqnarray}\label{approximate_para}
 D &\simeq&
 \frac{2+4(F/2M)^{\frac{3}{2}}(M_{p}\eta)^{-2}}
 {[1+2(F/2M)^{\frac{3}{4}}(M_{p}\eta)^{-1}]^{2}}~,\nonumber\\
 c_s^2 &\simeq& \frac{1+2(F/2M)^{\frac{3}{4}}(M_{p}\eta)^{-1}
 -2(F/2M)^{\frac{3}{2}}(M_{p}\eta)^{-2}}{3[1
 +2(F/2M)^{\frac{3}{2}}(M_{p}\eta)^{-2}]}~.
\end{eqnarray}

In the far past when $|\eta| \sim \sqrt{t_0-t} \gg 1$, and
considering only the leading order terms, we obtain the explicit
result
\begin{eqnarray}
 D \simeq 2~, ~~c_s^2 \simeq \frac{1}{3}~.
\end{eqnarray}
In this case, Eq. (\ref{eompert}) describes a massless oscillator,
and its solution is exactly
\begin{eqnarray}\label{subhorizon}
 u_k(\eta) \simeq \frac{e^{ic_sk\eta}}{\sqrt{2c_sk}}~.
\end{eqnarray}
Note that, during nearly all of the contracting phase our universe
is radiation-dominated, with a background EoS of approximately $1/3$
(recall Fig.~\ref{wfig}). Therefore the semi-analytic result
(\ref{approximate_para}) is valid before the bouncing phase, and
will not be much sensitive to the parameters $F$ and $M$, since the
terms containing these parameters are all negligible in the far
past, as can be seen from Eq. (\ref{approximate_para}). Thus, the
perturbation variable will always oscillate since its effective mass
squared vanishes. This behavior does not stop until the quintom
scenario takes place and then the bouncing phase occurs. In this
case, the cosmological perturbations seeded by the Galileon field in
the primordial era are stable and form a blue spectrum as in the
pure Galilean Genesis model. Obviously, this is incompatible with
current observations.

In order to produce a scale-invariant primordial power spectrum in a
Galileon bounce model, we are forced to consider the non-adiabatic
modes of the cosmological perturbations. One possible mechanism,
which was applied to a similar problem in
Ref.~\cite{Creminelli:2010ba}, is to introduce a scalar field which
plays the role of a curvaton in bouncing cosmologies. Provided that
this curvaton field couples to the Galileon field in a suitable
matter, we expect that its fluctuations will give rise to a
scale-invariant power spectrum. We will discuss this issue in the
next section. Before that, we turn our attention to tensor
perturbations.

\subsection{Tensor perturbations}

At leading order, the perturbed metric with respect to tensor
perturbations is
\begin{eqnarray}\label{metrictensor}
 ds^{2} = -a^2(\eta) [d\eta^2 -(\delta_{ij}+\gamma_{ij})dx^{i}dx^j]~
\end{eqnarray}
where $\gamma_{ij}(x)$ is the tensor perturbation. Since in the
Galileon model the higher derivative term of the Galileon field does
not play a role in tensor perturbations \cite{Gao:2011qe}, we can
write down the following second order action for tensor
perturbations
\begin{eqnarray}\label{actiontensor}
{\cal S}_2^T = \frac{1}{2}\int d\eta d^3x a^2 [\gamma_{ij}^{\prime2}
-(\partial_{k}\gamma_{ij})^2]~.
\end{eqnarray}
By imposing the traceless and transverse conditions on
$\gamma_{ij}$, its 9 components contain only two degrees of freedom.
These may be expanded as usual \cite{Stewart:1993bc}:
\begin{eqnarray}
 \gamma_{ij} = \sum_{\lambda=1}^2 \int
 \frac{d^3k}{(2\pi)^{3/2}}\gamma_{k,\lambda}e^\lambda_{ij}e^{ikx}~.
\end{eqnarray}
After defining $v^\lambda_g(k) \equiv a \gamma_{k,\lambda}$, we
obtain the equation of motion for the variable $v^\lambda_g(k)$
\begin{eqnarray}\label{eomtensor}
 v_{g}^{\lambda}(k)''+(k^{2}-\frac{a''}{a})v^\lambda_{g}(k)=0~.
\end{eqnarray}

Since in the contracting phase the universe is radiation-dominated
and thus $a(\eta)\sim\eta$, the last term in the above equation
vanishes. This means, like the case of scalar cosmological
perturbations, that there is only a massless oscillating solution:
\begin{eqnarray}
 v^\lambda_{g}(k) = \frac{e^{ik\eta}}{\sqrt{2k}}~.
\end{eqnarray}

However, the tensor perturbation will always get frozen at a
super-Hubble scale after the nonsingular bounce. In that case, from
the above calculation, right after the expansion of the universe
starts, the tensor perturbation will give rise to a blue-tilted
spectrum:
\begin{eqnarray}
 {\cal P}^T(k) \equiv \frac{k^3}{2\pi^2}|v^\lambda_{g}(k)|^2 \sim k^2~,
\end{eqnarray}
with spectral index
\begin{eqnarray}
 n_T \equiv \frac{d\ln{\cal P}^T(k)}{d\ln{k}}=2~.
\end{eqnarray}
Thus, for small $k$ modes within observable scales, the amplitude of
the tensor spectrum is severely suppressed. Correspondingly, the
scalar-to-tensor ratio $r \equiv {\cal P}^T/{\cal P}^S$ is expected
to be extremely small, and so the tensor perturbations of Galileon
bounce cosmologies will be very difficult to be detect in future CMB
observations.

\section{The Bounce Curvaton}

A bounce curvaton scenario can be employed to generate a primordial
power spectrum which is consistent with CMB observations
\cite{Cai:2011zx}.  It is known that entropy fluctuations are very
important for the formation of the primordial power spectrum. In the
context of inflationary cosmology, it has been realized that entropy
fluctuations can lead to an additional source of curvature
perturbations. This is now known as the curvaton mechanism. Entropy
fluctuations also play an important role in the Ekpyrotic scenario
\cite{Buchbinder:2007ad, Creminelli:2007aq}. In that scenario, the
primordial adiabatic fluctuations in the contracting phase acquire a
deep blue spectrum, whereas the entropy modes induced by light
scalar fields will be scale-invariant. Our model is similar to the
case of Ekpyrotic scenario in that the adiabatic fluctuations seeded
by the Galileon give rise to a blue spectrum. Thus, we expect that
the bounce curvaton mechanism can contribute to the curvature
perturbations so as to produce a scale-invariant power spectrum.

\subsection{The bounce curvaton scenario with kinetic coupling to
the Galileon}

We consider a massless curvaton field $\sigma$ kinetically coupled
to the background Galileon
\begin{eqnarray}\label{action_sigma}
 {\cal S}_{\sigma} = \int d^4x \sqrt{-g} [-\Pi^q(\partial\sigma)^2]~,
\end{eqnarray}
where $q$ will be determined momentarily. A similar curvaton
scenario was studied in Refs. \cite{Antoniadis:1996dj,
Rubakov:2009np, Hinterbichler:2011qk, Libanov:2011zy}. A
conformally-invariant coupling was considered and it was argued that
a strong-coupling regime can be avoided  \cite{Hinterbichler:2011qk,
Libanov:2011zy}. We would like to generalize this type of coupling
term. We note that this coupling, like the coupling of the Galileon
to gravity, never preserves the Galilean conformal
symmetry\footnote{The exponential coupling which preserves this
symmetry yields a scale-invariant spectrum in the usual Galilean
Genesis cosmology which is Minkowski in the infinite past, but not
our cosmology, which begins in a radiation-dominated phase.  This is
because substituting our radiation-dominated contracting solution
into the effective inverse metric $\sqrt{g}g^{\mu\nu}e^{2\Pi}$ does
not yield a fake de Sitter metric, however the $\Pi^{-4}$ coupling
to be described below does have this property.}.

Working with conformal time, the equation of motion for a mode of
the curvaton field with a fixed comoving wave number $k$ takes the
form
\begin{eqnarray}
 \sigma_k'' + (2{\cal H}+\frac{q\Pi'}{\Pi})\sigma_k' + k^2\sigma_k =0
 ~,
\end{eqnarray}
where ${\cal H}=aH$ is the conformal Hubble parameter. From the
analysis of background solution, we already know that the universe
is initially radiation-dominated in the contracting phase, which
implies that $\Pi \sim a \sim \eta$. Thus, we have ${\cal H} \simeq
1/\eta$, and so the equation of motion simplifies to
\begin{eqnarray}\label{eomsigma}
 \sigma_k'' + \frac{2+q}{\eta}\sigma_k' + k^2\sigma_k = 0~.
\end{eqnarray}
The general solution to this equation can be expressed as a
combination of two Bessel functions
\begin{eqnarray}\label{solution_sigma}
 \sigma_k = |\eta|^{-\frac{1+q}{2}} \bigg[
 c_1J_{-\frac{1+q}{2}}(k|\eta|)
 +c_2J_{\frac{1+q}{2}}(k|\eta|) \bigg] \ .
\end{eqnarray}
Here $c_1$ and $c_2$ are two coefficients which may in principle
depend on $k$. However, when we match the asymptotic behavior of the
solution (\ref{solution_sigma}) in the limit of $k|\eta|\gg 1$ with
the sub-Hubble solution $\sigma_k\rightarrow e^{ik\eta}/\sqrt{2k}$,
we find that both $c_1$ and $c_2$ are independent of the comoving
wave number $k$. Consequently, the asymptotic behavior of
(\ref{solution_sigma}) at super-Hubble scales is
\begin{eqnarray}\label{superhorizon}
 \sigma_k \sim \tilde{c}_1k^{-\frac{1+q}{2}}|\eta|^{-(1+q)}
 +\tilde{c}_2k^{\frac{1+q}{2}}~,
\end{eqnarray}
which consists of a constant mode and a varying (either decaying or
growing) mode. Here $\tilde{c}_i$ and $c_i$ differ by an irrelevant
factor.

\subsection{Scale-invariant power spectrum}

We consider two cases:\\
i) $q=2:$ In this case, the super-Hubble solution for $\sigma_k$
becomes
\begin{eqnarray}
 \sigma_k \sim \tilde{c}_1k^{-\frac{3}{2}}|\eta|^{-3}
 +\tilde{c}_2k^{\frac{3}{2}}~.
\end{eqnarray}
Since in the radiation-dominated contracting phase $|\eta|$ is a
decreasing function with respect to cosmic time $t$, the first term
of this solution is growing and thus will be dominant. Therefore the
power-spectrum ${\cal P}_\sigma$ will be
\begin{eqnarray}
 {\cal P}_\sigma &\equiv& \frac{k^3}{2\pi^2}|\sigma|^2 \nonumber\\
  &\simeq& \frac{\tilde{c}_1^2}{2\pi^2}|\eta|^{-6}~,
\end{eqnarray}
which is scale invariant and its amplitude is growing.\\
ii) $q=-4:$ In this case, the super-Hubble solution of $\sigma_k$ is
\begin{eqnarray}
 \sigma_k \sim \tilde{c}_1k^{\frac{3}{2}}|\eta|^{3}
 +\tilde{c}_2k^{-\frac{3}{2}}~.
\end{eqnarray}
For the same reason as above, here the first term is a decaying mode
and thus the dominant mode is the last term. Thus, the power
spectrum is
\begin{eqnarray}
 {\cal P}_\sigma \simeq \frac{\tilde{c}_2^2}{2\pi^2}~,
\end{eqnarray}
which is also scale invariant and its amplitude is a constant.


In order to determine whether the curvaton mechanism is reliable,
one has to estimate the backreaction of the curvaton field on
background dynamics. From the action (\ref{action_sigma}), one can
easily find the effective energy density of the curvaton field $
\rho_{\sigma} = \Pi^q \dot\sigma^2 $. Solving the equation of motion
(\ref{eomsigma}) in the contracting phase, we obtain $\sigma \sim
(t_0-t)^{-\frac{1+q}{2}}$. Thus, the effective energy density of
$\sigma$ takes the form,
\begin{eqnarray}
 \rho_\sigma \sim (t_0-t)^{-3-\frac{q}{2}}~,
\end{eqnarray}
where we have applied the solution of $\Pi$ in the
radiation-dominated contracting phase. Recall that, since the
universe shrinks in a radiation-dominated phase, the background
energy density evolves as $\rho \sim (t_0-t)^{-2}$. Therefore, a
stable evolution for the curvaton field requires $q \geq -2$. As a
consequence, we find the case of $q=2$ is a stable solution for the
bounce curvaton scenario.   However, in the case $q=-4$, although
one obtains a scale-invariant spectrum of isocurvature perturbation,
one needs to fine tune the initial condition of the curvaton field
very carefully to prevent its backreaction from destroying the
Galileon bounce.

Note that the fluctuations seeded by the curvaton field $\sigma$ are
mainly isocurvature modes in the contracting phase. However, they
may easily be converted into curvature perturbations after the
bounce. In this process, there are two important effects. First, for
the adiabatic modes of cosmological perturbations in the contracting
phase, the corresponding spectrum is blue tilted so that its
amplitude is strongly suppressed at large length scales. Thus the
main contribution to curvature perturbations at large length scales
(corresponding to the observable scales) automatically come from the
non-suppressed fluctuations of curvaton field. Second, after the
bounce, it is expected that there exists a preheating phase during
which the curvaton plays the role of an entropy field and reheats
the universe \cite{Levasseur:2011mw} (see also \cite{Feng:2002nb}
for an earlier study of the scenario of curvaton reheating). In this
phase, the entropy fluctuations of the curvaton are transformed into
curvature perturbations at all length scales.

\section{Conclusions and Discussion}

In this paper we have presented and studied bouncing solutions of
the conformal Galileon model. The Galileon effective field
description shows that the NEC can be violated safely without the
usual ghost and gradient instabilities. Moreover, the terms in the
Lagrangian leading to the violation of NEC are higher order
operators, which are typically suppressed at low energy scales. Thus
it is natural to consider NEC violation at high energy scales in a
Galileon model, and if applied to cosmology, NEC violation usually
occurs in the early universe.  We study the conformal Galileon as a
toy model. We find that in this model the universe can begin in a
radiation-dominated contracting phase. Along with the contraction,
the EoS of the universe evolves from $w=1/3$ to a regime of Phantom
phase $w<-1$ and thus a quintom scenario is realized. After that,
the universe passes through a nonsingular bouncing phase and finally
approaches the moment of Galilean Genesis. We examined this solution
both semi-analytically and numerically.

After having analyzed the background dynamics, we studied the
perturbation theory of this model. It is well-known that, for a
generic cosmological model described by a single k-essence field,
the cosmological perturbations diverge in quintom scenarios
\cite{Xia:2007km}. In the past people were forced to consider models
involving multiple degrees of freedom to circumvent this problem.
However, in the Galileon model, although the number of degrees of
freedom does not change, the stress energy tensor is in general that
of an imperfect fluid in the sense that once one includes
perturbations the viscosity is nonzero  \cite{fluid}. The novel
behavior of this fluid allows it to avoid the usual divergence of
perturbations when the EoS of the model crosses over the
cosmological constant boundary. Our result confirms this conclusion
and also shows that, for linear perturbations about Galileon
bouncing cosmologies, there is no ghost instability and no
superluminal propagation for linear perturbation modes. Therefore,
in this background, the model is well-behaved and causal. However,
our analysis shows that the adiabatic modes of cosmological
perturbations yield a blue tilted spectrum, and so do tensor
perturbations. In order to be consistent with the current
cosmological observations, we include a  bounce curvaton which
couples to the Galileon field through its kinetic term. We obtained
two such possible couplings which yield a scale-invariant power
spectrum, with one possessing a growing spectrum in the contracting
phase and the other a constant spectrum. The coupling to curvaton in
both cases, however, breaks conformal symmetry explicitly. We
examined the back reaction of curvaton on the background evolution,
and find that in one case the bouncing solution is safe from the
effect of curvaton, while in the other case fine-tuning might be
needed.

At the end of the paper, we would like to comment on some issues.
First, we ought to be aware of the final state of this model, i.e.,
the Galilean Genesis. A promising approach to avoiding strong
coupling is to introduce a phase of preheating right before the
moment of Galilean Genesis, as analyzed in Ref.
\cite{Levasseur:2011mw}. Although the authors of
\cite{Levasseur:2011mw} performed their analysis in the background
of an emergent universe, their results can be roughly applied to
bouncing cosmologies, providing an important supplement to the
bounce preheating scenario \cite{Cai:2011ci}. Another interesting
issue relates to the predictions of a Galileon bounce. From the
study of linear perturbations, we predict that as in the original
Galilean genesis proposal, primordial tensor perturbations will have
a blue spectrum.  This means that it will be difficult to observe
the tensor spectrum in future CMB experiments. Obviously this is not
a good news for this Galileon model of cosmology. However, we might
have chance to provide experimental signatures for a Galileon bounce
if we study nonlinear perturbations. Namely, it is known that, for a
bouncing universe with an EoS of order unity in the contracting
phase, the corresponding non-Gaussianities are quite sizable and
might be sensitive to the forthcoming cosmological data
\cite{Cai:2009fn}. Therefore, these non-Gaussianities in
cosmological perturbations merit further study in a future project.

\begin{acknowledgments}
We are pleased to thank R. Brandenberger, D. Easson, Xian Gao and A.
Vikman for discussions and comments. TQ is funded in part by the
National Science Council of R.O.C. under Grant No.
NSC99-2112-M-033-005-MY3 and No. NSC99-2811-M-033-008 and by the
National Center for Theoretical Sciences. JE is supported by the
Chinese Academy of Sciences Fellowship for Young International
Scientists grant number 2010Y2JA01. YC is supported by funds of
department of physics at Arizona State University. ML is supported
by the NSFC under Grants Nos. 11075074 and 11065004 and the
Specialized Research Fund for the Doctoral Program of Higher
Education under Grant No. 20090091120054. The research of XZ is
supported in part by the National Science Foundation of China under
Grants No. 10533010 and 10675136, by the 973 program No.
2007CB815401, by NSFC No. 10821063, and by the Chinese Academy of
Sciences under Grant No. KJCX3-SYW-N2. YC wish to thank Professor
Xinmin Zhang and the Theory Division of the Institute for High
Energy Physics for hospitality during the period when the current
project was initiated.
\end{acknowledgments}
\vspace{.5cm}

{\it Note added}-After completing this manuscript we learned that a
treatment of bouncing solutions in a more general set of theories
has been done by Easson, Sawicki and Vikman, which will soon appear
in Ref.~\cite{Easson:2011zy}.

\end{document}